\documentclass[reprint,american,aps,pre,superscriptaddress,showpacs,showkeys,floatfix]{revtex4-2}
\usepackage{bm}  
\usepackage{graphicx}
\usepackage{amsmath}
\usepackage{enumitem}
\usepackage{amsfonts}
\usepackage{amssymb}
\usepackage{hyperref}

\begin{document}

\title{Stopping Power Enhancement From Discrete Particle-Wake Correlations in High Energy Density Plasmas}

\author{I. N. Ellis}
\email{iellis@minevaluation.com}
\affiliation{Lawrence Livermore National Laboratory, Livermore, California 94550, USA}
\affiliation{University of California, Los Angeles, California 90095, USA}

\author{D. J. Strozzi}
\affiliation{Lawrence Livermore National Laboratory, Livermore, California 94550, USA}

\author{W. B. Mori}
\author{F. Li}
\affiliation{University of California, Los Angeles, California 90095, USA}

\author{F. R. Graziani}
\affiliation{Lawrence Livermore National Laboratory, Livermore, California 94550, USA}

\date{\today}

\begin{abstract}
Three-dimensional (3D) simulations of electron beams propagating in high energy density (HED) plasmas using the quasi-static Particle-in-Cell (PIC) code QuickPIC demonstrate a significant increase in stopping power when beam electrons mutually interact via their wakes.  Each beam electron excites a plasma wave wake of wavelength $\sim2\pi c/\omega_{pe}$, where $c$ is the speed of light and $\omega_{pe}$ is the background plasma frequency. We show that a discrete collection of electrons undergoes a beam-plasma like instability caused by mutual particle-wake interactions that causes electrons to bunch in the beam, even for beam densities $n_b$ for which fluid theory breaks down.  This bunching enhances the beam's stopping power, which we call ``correlated stopping,'' and the effect increases  with the ``correlation number'' $N_b \equiv n_b (c/\omega_{pe})^3$.  For example, a beam of monoenergetic 9.7 MeV electrons with $N_b=1/8$, in a cold background  plasma with $n_e=10^{26}$ cm$^{-3}$ (450 g cm$^{-3}$ DT), has a stopping power of $2.28\pm0.04$ times the single-electron value, which increases to $1220\pm5$ for $N_b=64$.  The beam also experiences transverse filamentation, which eventually limits the stopping enhancement.
\end{abstract}


\maketitle

Energetic particle stopping power is a critical issue in many plasma physics contexts, including self-heating by fusion products, magnetic fusion devices, space plasmas, cancer therapy, and high energy density (HED) systems.  We focus on the last, where energetic (non-thermal) charged particles are of interest for several reasons.  Laser-plasma interactions, such as stimulated Raman scattering and two-plasmon decay, produce energetic electrons that alter energy coupling in an inertial fusion system.  Ultra-intense short-pulse lasers can also produce energetic ions and relativistic electrons.  An interesting application is the Fast Ignition (FI) approach to inertial fusion \cite{tabak:ignition}, in which a beam of energetic electrons (ideally with kinetic energy $\sim1-3$ MeV) deposits energy into a compressed target's core, beginning the ignition process.  This application motivates our choice of parameters.  The important role of electron wakes in this work is also relevant to plasma-based particle accelerator research.

Most calculations of electron beam transport for MeV and higher particle energies in HED plasmas use a single-electron stopping formula based on quantum electrodynamics (QED) and a collective dielectric response or wake (discussed in Appendix \ref{Appendix-A}) \cite{icru37, Deutsch:Interaction, solodov:stopping, atzeni:stopping, Robinson:Theory}:
\begin{align}
  \frac{d\gamma}{ds} & = -\frac{e_b^2}{m_bc^2} \frac{\omega_{pe}^2}{v^2} \ln\Lambda^{qm}; \nonumber\\
  \Lambda^{qm} & \equiv [2(\gamma-1)]^{1/2} \frac{2\pi\delta_{sk}}{\lambda_{db}}.
  \label{QED-dEdx}
\end{align}
We use CGS units throughout.  Energy loss per distance traveled $s$ is to background electrons, and $-e$, $m_e$, $n_e$, and $\omega_{pe}\equiv(4\pi n_ee^2/m_e)^{1/2}$ are the electron charge, mass, number density, and plasma frequency respectively, and $\delta_{sk} \equiv c/\omega_{pe}$ is the collisionless skin depth. We have omitted small non-logarithmic terms and radiative loss, the latter of which is small in hydrogen for electron energies $<100$ MeV, though for high-Z materials like gold it is significant for $\sim10$ MeV \cite{icru37, ESTAR}.  The beam electron has charge $e_b$ (distinguished from $-e$ to show correlation effects), mass $m_b$, speed $v$ and $\beta=v/c$, Lorentz factor $\gamma=[1-\beta^2]^{-1/2}$, kinetic energy $E=m_bc^2(\gamma-1)$, and de Broglie wavelength $\lambda_{db}\equiv h/m_ev$.  Eq.\ \ref{QED-dEdx} applies for $E\gg T_e$, with $T_e$ the background electron temperature, and $\Lambda^{qm}$ assumes $m_b=m_e$, $e_b=-e$. The stopping power in Eq.\ \ref{QED-dEdx} scales as $e_b^2/m_b$, so if $N$ beam electrons act like a single ``macro-particle" with $e_b=-Ne$, $m_b=Nm_e$, then their stopping is $\propto e_b^2/m_b\propto N$ \cite{Bret:Correlated}.  This is the basic idea behind correlated stopping, and requires discrete beam particles -- it would not occur for a smooth beam of ``jellium."  Such stopping has also been studied in the field of electron plasma accelerators in an attempt to design a more compact beam dump \cite{Wu:Collective}.

In this paper, we closely examine how the stopping power of a collection of particles can be enhanced above the single-particle stopping, Eq.\ \ref{QED-dEdx}, due to ``correlated stopping," in which the beam electrons mutually interact via their plasma wave wakes.  An increase in stopping power may increase the energy deposited by an electron beam in an FI target core, thus making the concept more feasible.  In contrast to the ``collective stopping" considered by others \cite{Malkin:Collective}, in which fluid beam-plasma instabilities lead to an increase in stopping power, correlated stopping is caused by discrete particle-wake interactions. This can occur when the electron beam density $n_b$ is too low for a fluid description to strictly apply, such as in a FI target core.  Collective stopping has been studied extensively for ion beam stopping \cite{Arista:Energy,*Arista:Stopping,*Bringa:Collective,*Bringa:Energy,*Miskovic:Dynamics} and for static beam electron configurations \cite{Thomas:Stopping, Deutsch:Correlated, Bret:Correlated}.  We present here the first 3D PIC simulations of dynamic correlated electron stopping in HED plasmas.  We observe a consistent increase in stopping power beyond the single-particle result with increasing ``correlation number'' $N_b$ defined as:
\begin{equation}
N_b \equiv n_b\delta_{sk}^3 = 4.50\times 10^6 \frac{n_b}{n_c} \left(\frac{n_c}{n_e}\right)^{3/2} \lambda[\mu\mathrm{m}].
\end{equation} 
The second practical form is for electron beams produced by a short-pulse laser of wavelength $\lambda$ with $n_c$ the laser critical density.

\section{PIC method}
We examine the stopping power for finite-size relativistic electrons using the particle-in-cell (PIC) method. The standard PIC method can suffer from numerical \v{C}erenkov radiation as well as self-forces created by aliasing. Numerical \v{C}erenkov radiation \cite{greenwood:elimination, Xu:recent}, which is a common issue in finite-difference (FD) electromagnetic PIC codes, is caused by particles moving faster than light propagates on the mesh. Issues related to numerical \v{C}erenkov radiation are different than those that arise from the numerical \v{C}erenkov instability \cite{Godfrey:Numerical, *Godfrey:Canonical, *Yu:Modeling, *Godfrey:Numerical2013, *Xu:Numerical}. Aliasing can also lead to artificial self-fields on the particle, even for solvers with superluminal light waves or perfect dispersion solvers. These effects can increase the single-electron stopping power. New solvers developed for FD PIC codes may mitigate numerical \v{C}erenkov radiation \cite{Lehe:Numerical, *Fei:Controlling, Xu:recent} and permit studies of correlated stopping in divergent beams in the future.

To circumvent these issues, we use the quasi-static PIC code QuickPIC \cite{Huang:QuickPIC, An:Improved}. QuickPIC uses coordinates $(x, y, \xi \equiv ct-z, s\equiv z)$, where $z$ is the direction of beam propagation.  The quasi-static approximation is $\partial/\partial s \ll \partial/\partial \xi$, meaning the length-scale of variations of the beam or wake with $s$ is much greater than the wake wavelength; i.e., the beam evolves on a time-scale much slower than it takes a beam particle to pass a plasma particle.  This approximation decouples the wake calculation from the beam particle push, and allows much larger time-steps than fully electromagnetic codes.  QuickPIC does not include radiative fields, has similarities to the Darwin approximation, and is \textit{not} an electrostatic model. QuickPIC sends a 2D plasma slice across the box in the $\xi$-direction at each $s$-step.  We can therefore view $\xi$ as the ``time'' after the box begins passing through a transverse plasma slice at position $s$.


\begin{table}[htbp]
\centering
\begin{tabular}{ |l|l| }
  \hline
  \multicolumn{2}{|c|}{QuickPIC Simulation Parameters} \\
  \hline
  $n_e$ & $10^{26}$ cm${}^{-3}$ \\
  $T_e$ & 0 eV \\
  Interpolation & Linear\\
  Cell Width & $\Delta_x = \Delta_y = \Delta_\xi = 0.0405 \delta_{sk}$\\
  $ds$ & 2$\delta_{sk}$\\
  Small Box Dimensions & $41.515\delta_{sk}\times[1,1,2]$ in $[x, y, \xi]$\\
  Small Beam Dimensions & $10\delta_{sk}\times[1,1,8]$ \\
  Large Box Dimensions & $83.03\delta_{sk}\times[1,1,2]$\\
  Large Beam Dimensions & $40\delta_{sk}\times[1,1,8]$ \\
  \hline
\end{tabular}
\caption{The parameters for the QuickPIC simulations.\label{table-qp-sim-params}}
\end{table}

The simulation parameters are listed in Table \ref{table-qp-sim-params}.  They are relevant to an electron beam propagating through a fully-ionized deuterium-tritium (equal atomic fraction) plasma of $\approx450$ g cm$^{-3}$, where the background electron number density $n_e=10^{26}$ cm$^{-3}$; typical of the compressed fuel in FI designs. Under these conditions, if $n_b$ is the  critical density $n_c\approx 1.11 \times 10^{21}$ cm$^{-3}$ for $1\ \mu$m light, then $N_b=0.15$. For these conditions, $\ln\Lambda^{qm}=8.35$.

We primarily study monoenergetic beams with $N_b$ from 1/8 to 64 and momentum $p_z=20m_ec$ ($E=9.7$ MeV).  The energy, while larger than the $\sim1-3$ MeV in an optimal FI beam \cite{atzeni:stopping, Strozzi:Fast, Robinson:Theory}, is chosen to ensure the validity of the quasi-static approximation.  We expect the instabilities we observe in our simulations to evolve faster in beams closer to the 1-3 MeV range due to the lower Lorentz factor.

Two major assumptions we make are using a cold background plasma (plasma electrons are initialized with zero initial velocity) and neglecting collisions.  Since the stopping power is independent of $T_e$ for $E\gg T_e$, we do not expect our results to vary appreciably with finite $T_e$.  The question of how $T_e$ could affect the plasma wakes is future work, though a brief discussion is provided in Appendix \ref{Appendix-B}, and we note that the plasma-based accelerator literature has used Green's functions for $T_e=0$ for many years.  Finite $T_e$ also leads to a numerical instability that develops in the particle wake when a warm plasma is used. The increased temperature will cause transverse spreading of the wake via diffraction \cite{fahlen:dissertation} and will affect the motion of the beam particles via discrete particle thermal fluctuations.

As for collisions, the background electron-ion collision frequency for momentum transfer, including Fermi degeneracy, is $\nu \approx 7.46\times10^{-6} n_e [(n_e/2.05\times10^{22})^{2p/3} + T_e^p]^{-3/2p}$, $p=1.72$, ($n_e$ in cm$^{-3}$, $T_e$ in eV, $\nu$ in s$^{-1}$) \cite{Robinson:Theory}.  When $n_e=10^{26}$ cm$^{-3}$, $\omega_{pe}\approx5.64\times10^{17}$ rad/s. At $T_e=0$ eV, $\nu/\omega_{pe} \approx 0.27$ and neglecting collisions is unrealistic. At $T_e=1$ keV, which is typical of the fuel in FI designs at the time the electron beam starts, $\nu/\omega_{pe} \approx0.038$ and the collisionless assumption is more feasible.  As the fuel heats, collisions become less important, e.g.\ at $T_e=5$ keV, $\nu/\omega_{pe} \approx 0.0037$.  Collisions should generally be less important in other HED systems at lower electron densities than the $10^{26}$ cm$^{-3}$ we consider.  Despite these limitations, our work with a cold plasma provides significant insight into correlated stopping.

We set up an electron beam centered transversely in the box.  We simulate two different beam sizes, whose dimensions and respective box sizes are listed in Table \ref{table-qp-sim-params}.  Both sizes have cell width $\Delta=0.0405\delta_{sk}$.  There is one background electron per cell and, for both the beam and plasma, one PIC particle represents one physical particle, avoiding the enhanced stopping experienced by macro-particles (a PIC particle typically has a charge and mass of many electrons) \cite{May:Enhanced}.  

The simulations use a window moving at $c$ in the direction of beam propagation $\hat{z}$.  The transverse boundaries are conducting with specular reflection for the particles.  The 2D plasma sheet is initialized with the plasma particles in a stable configuration.  Therefore, with a cold plasma, as the sheet crosses the box, the arrangement of the plasma particles does not change unless there is a beam particle present.

We perform simulations for four cases:
\begin{itemize}[noitemsep,nolistsep]
\item[1)] small, monoenergetic beam; immobile ions
\item[2)] same as 1) but mobile ions of charge $+e$ and mass $1836.15m_e$
\item[3)] same as 1) but 1 MeV beam temperature in $z$ (cold in transverse directions)
\item[4)] same as 1) but large beam
\end{itemize}
All electrons in the monoenergetic beams are initialized with a momentum $p_z=20m_ec$ ($E=9.7$ MeV).  When the beam has a temperature in $z$, the electrons are initialized using a Maxwell-J\"uttner distribution with $T_b=$ 1 MeV centered around $p_z=20m_ec$.  The assumption of no transverse temperature is unrealistic \cite{May:Mechanism}.  For each simulation case, we run with $N_b=1/8$, 1, 8, and 64 by varying $n_b$ and keeping $n_e=10^{26}$ cm$^{-3}$ fixed.  For example, when $N_b=64$, $n_b=4.26 \times 10^{23}$ cm$^{-3}$. For each $N_b$, we run eight simulations.  For each run, the beam electrons are initially placed on a cubic lattice of spacing $\Delta_l=\delta_{sk}/N_b^{1/3}$, then displaced in each Cartesian direction by a random distance chosen uniformly from $[-\Delta_l/2, \Delta_l/2)$.

\begin{figure}[tbp]
	\centering
	\includegraphics[width=\linewidth]{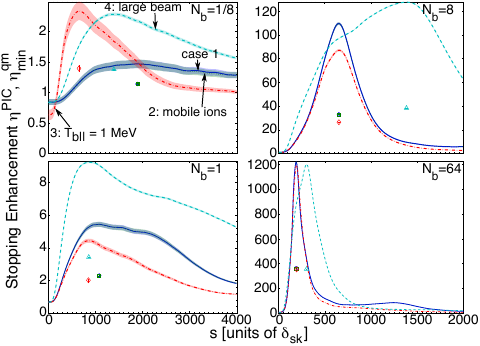}
    \caption{The $s$-evolution of the stopping power enhancement $\eta^{PIC}$ from Eq.\ \ref{eq:etapic} for four values of $N_b$.  Each line is an average over eight QuickPIC runs, and the associated transparent blooming is the associated uncertainty, which can become thinner than the line itself as $N_b$ increases.  We also plot the minimum physical stopping enhancement $\eta^{qm}_{min}$ from Eq.\ \ref{eq:etaqm} at the $s$ of maximum $\eta^{PIC}$.  The results with and without mobile plasma ions are mostly indistinguishable.\label{dEdx-plots}}
\end{figure}

\begin{figure}[tbp]
	\centering
    \includegraphics[width=\linewidth]{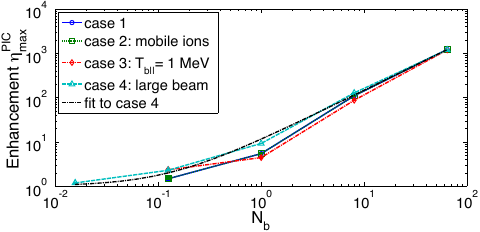}
    \caption{The peak stopping enhancement $\eta^{PIC}$ vs. $N_b$ for all four simulation cases.  A curve fit for the large box, case 4, is also plotted.\label{eta-vs-Nb}}
\end{figure}

\section{Correlated stopping results}
Fig.\ \ref{dEdx-plots} shows the $s$-evolution of the stopping power enhancement averaged over eight runs for each $N_b$ and simulation type.  We first find $d\gamma/ds|_N^{PIC}$, the stopping of one electron in a full, $N$-beam-electron simulation, by averaging over all beam electrons, up to 16,384,000 for $N_b=64$ in large-beam simulations, then average the eight results and find the standard deviation.  The stopping enhancement 
\begin{equation}
\eta^{PIC} \equiv \frac{d\gamma/ds|_N^{PIC}}{d\gamma/ds|_1^{PIC}}, \label{eq:etapic}
\end{equation}
where $d\gamma/ds|_1^{PIC}$ is the stopping power of a lone beam electron measured in a separate QuickPIC simulation. $\eta^{PIC}$ rapidly moves above unity in all cases and increases with $N_b$, reaching a dramatic enhancement of $\sim10^3$ for $N_b=64$.  For ease of comparison, Fig.\ \ref{eta-vs-Nb} plots the peak values of $\eta^{PIC}$ from Fig.\ \ref{dEdx-plots} vs.\ $N_b$, along with that for $N_b=1/64$ in simulation case 4.  We include a curve fit for case 4: $\eta^{PIC}=1+10.58N_b^{1.14}$.

We now estimate the enhancement of the physical, quantum-mechanical stopping power: $\eta^{qm} \equiv(d\gamma/ds|_N^{qm}) / (d\gamma/ds|_1^{qm})$.  The single-particle stopping in QuickPIC $d\gamma/ds|^{PIC}_1$ is well below the single-particle quantum result $d\gamma/ds|^{qm}_1$, as discussed in Appendix \ref{Appendix-A}.  Our simulations therefore do not show how much this ``unresolved stopping," $d\gamma/ds|^{qm}_1-d\gamma/ds|^{PIC}_1$, is enhanced by correlation effects.  A likely upper bound is to assume the unresolved stopping is enhanced by the same factor as the stopping resolved in the PIC code, or $\eta^{qm}_{max}=\eta^{PIC}$.  For a lower bound, we assume none of the unresolved stopping is enhanced: $d\gamma/ds|^{qm}_N -d\gamma/ds|^{PIC}_N = d\gamma/ds|^{qm}_1 - d\gamma/ds|^{PIC}_1$, or
\begin{equation}
\eta^{qm}_{min} =\frac{\ln\Lambda_1^{PIC}}{\ln\Lambda_1^{qm}} \left( \eta^{PIC}-1 \right)+1 \approx 0.38\eta^{PIC} + 0.62. \label{eq:etaqm}
\end{equation}
This lower bound still gives significant stopping enhancement, as shown in Fig.\ \ref{dEdx-plots} by the discrete symbols.

Fig.\ \ref{dEdx-plots} clearly shows stopping power increasing with $N_b$.  The different simulation cases change the evolution of the stopping power for each $N_b$.  Mobile plasma ions make the least difference, as results with and without mobile ions are mostly indistinguishable.  This small effect is explained by the relatively small ion density perturbation.  For $N_b=64$, when $s=100\delta_{sk}$, at the tail of the beam, $\max(\delta n_i)/\max(\delta n_e) \approx 0.025$, which is negligible in this context. 

In all cases except $N_b=1/8$, adding a 1 MeV beam temperature in $z$ causes the stopping power to peak at approximately the same time as the monoenergetic beam but at a lower value, then remain below it thereafter.  However, for $N_b=1/8$, the temperature causes the stopping power to peak earlier and at a higher level, then drop below that of the monoenergetic beam.  This discrepancy may be a result of the small number of particles, 1,000 when $N_b=1/8$, and may disappear with a larger beam.

In the simulations using the large box, the stopping power reaches a higher peak level than that of monoenergetic beams in the small box in the cases of $N_b=1/8$, 1, and 8, and the stopping power remains above those of the smaller beams thereafter.  In all three cases, the stopping power grows more rapidly early in $s$ than for the smaller monoenergetic beams.   For $N_b=64$, the large-box stopping power peaks later than for the small beam and stays above it until $s\approx 800\delta_{sk}$.  For $N_b=1/8$, the stopping enhancement in the large box is still near 2 at $s=3,000\delta_{sk}$, which may have a significant impact on applications like FI.

\section{Beam-plasma-like instabilities and saturation}
\begin{figure}[tbp]
	\centering
	\includegraphics[width=\linewidth]{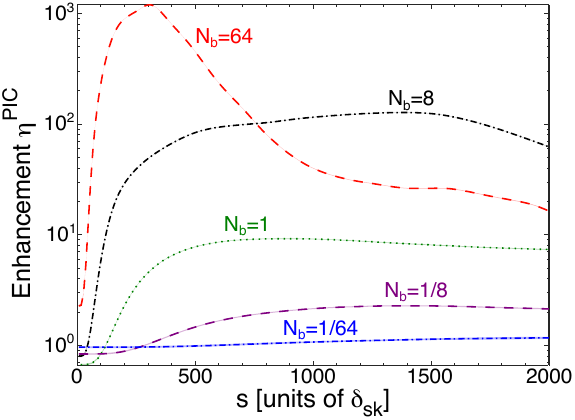}
    \caption{Stopping power enhancement for five values of $N_b$ using large beams (case 4), each averaged over eight runs.  The dashed curves indicate the average over the runs, and the thickness of the faint solid curve associated with each line is the uncertainty.  The faint solid curves are typically thinner than the dashed lines on the plot, and are clearer in Fig. \ref{dEdx-plots}. The relative uncertainty increases as the stopping power decreases, illustrating the increasing effect of particle discreteness with decreasing $N_b$.\label{dEdx-8avg}}
\end{figure}

The stopping power in all cases initially increases due to fluid-like instabilities, then peaks and begins to decrease due to saturation.  We say ``fluid-like" because, for the parameters used here, the inter-particle spacing can be $\sim$ the wake wavelength $2\pi\delta_{sk}$, and larger than the wake transverse radius $\delta_{sk}$, violating the continuum assumption of the fluid approximation.  The uncertainties in Fig.\ \ref{dEdx-plots} and the variation between runs in Fig.\ \ref{dEdx-8avg} illustrate the increasing effect of particle discreteness with decreasing $N_b$.  The spreading of a single-particle wake with increasing $\xi$, discussed in Ref.\ \cite{ellis:dissertation}, will also work to invalidate fluid results with decreasing $N_b$.

\begin{figure}[tbp]
	\centering
    \includegraphics[width=\linewidth]{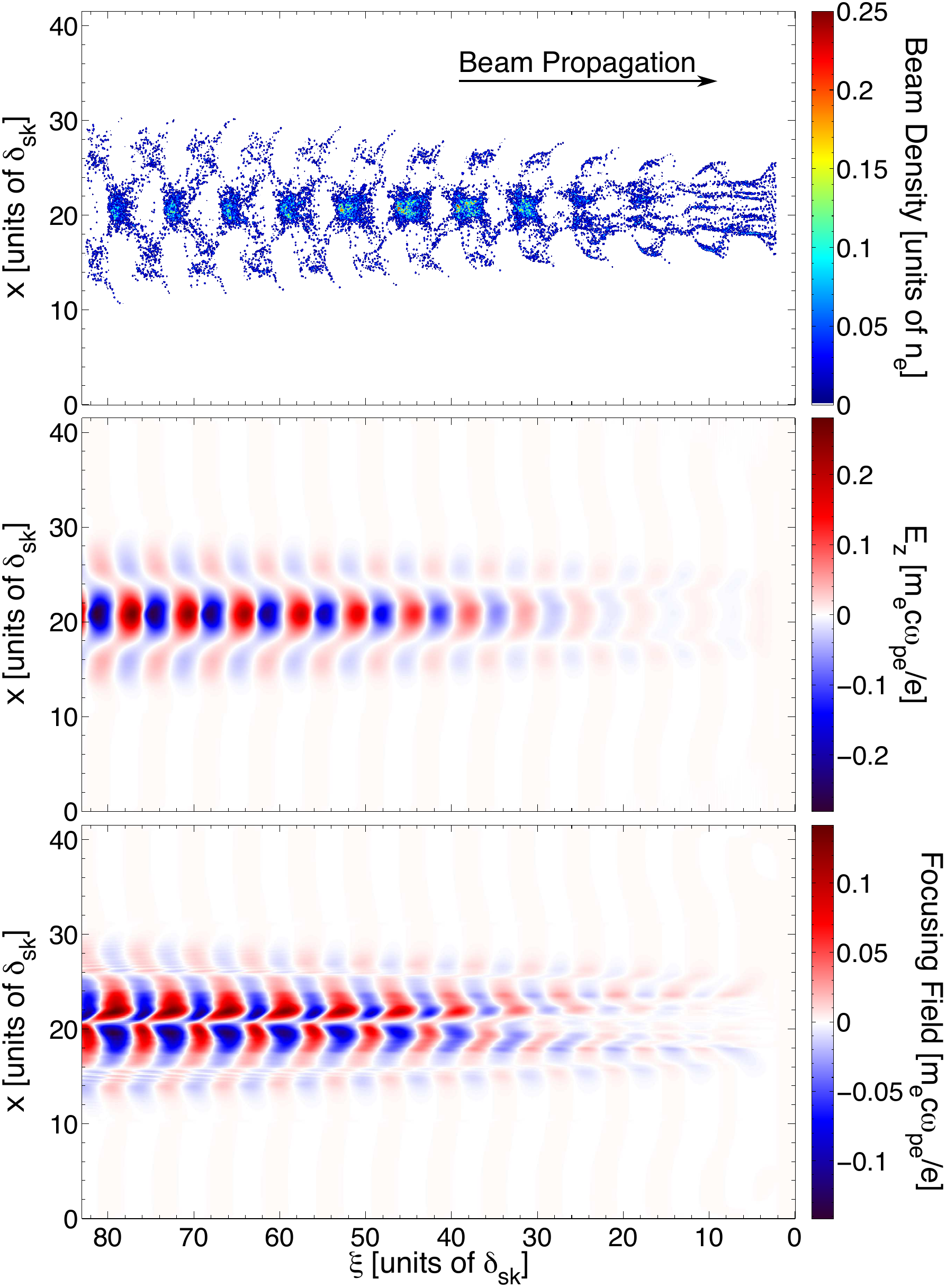}
    \caption{The beam density in a cut plane through the middle of the box at $y=20.76\delta_{sk}$ is plotted along with its corresponding longitudinal and focusing fields for simulation case 1 with $N_b=64$ at $s=200\delta_{sk}$.\label{qeb-fields-mono}}
\end{figure}

Fig.\ \ref{qeb-fields-mono} illustrates the effect of the fluid-like instabilities on a small monoenergetic beam with $N_b=64$ at $s=200\delta_{sk}$.  The beam contains regions of alternating bunching and spreading in all three dimensions, which is the primary source of the stopping enhancement.  This bunching is caused by the oscillating electric fields of the particle wakes in the longitudinal direction and the corresponding transverse focusing fields, as seen in the figure.  The longitudinal behavior is related to the fluid beam-plasma or two-stream instability \cite{Bret:Multidimensional}, and the transverse behavior is related to the transverse self-modulation instability \cite{Vieira:Transverse, *An:dissertation} studied in plasma wakefield accelerators.  As the particles begin to bunch, bunches tend to align in the logintudinal direction and merge in the transverse direction, with the transverse merging limited by beam size or filamentation.  Due to constructive interference, the wakefields are largest at the tail of the beam, and the process occurs most rapidly there.

Multidimensional electron beam-plasma instabilities have recently been studied in the relativistic regime \cite{Bret:Multidimensional}, and an exact kinetic theory for them has been developed using Maxwell-J\"uttner distribution functions \cite{Bret:Exact}.  The beam-plasma and transverse self-modulation instabilities in particular have been studied extensively in the context of laser-plasma interactions \cite{Mori:Physics}.  Ref.\ \cite{ellis:dissertation} has detailed derivations of them, the former of which is generalizable to 1D fluid streaming instabilities.  To briefly summarize, beam and plasma densities satisfy
\begin{align}
\left(\frac{\partial^2}{\partial s^2}+ \frac{k_{pb}^2}{\gamma_b^3}\right)\delta n_b &= -\frac{k_{pb}^2}{\gamma_b^3}\delta n_e, \nonumber \\
\left(\frac{\partial^2}{\partial \xi^2} + k_{pe}^2\right)\delta n_e &= -k_{pe}^2\delta n_b,\label{eq-instability}
\end{align}
with subscript $j=(b,e)$ for (beam, background plasma) quantities, $\delta n_j$ is the density perturbation, $k_{pj}\equiv\omega_{pj}/c$, and $(\omega_{pb},\gamma_b)$ are the beam (plasma frequency, Lorentz factor). We Fourier analyze with $\delta n_j\propto\exp[i(k_{\xi}\xi-k_s s)]$. Recall that $s$ and $\xi$ are akin to time and space, so instability entails $\mathrm{Im}[k_s]>0$ for real $k_{\xi}$. The unstable modes satisfy
\begin{equation}
\mathrm{Im}[k_s]=\frac{k_{pb}k_\xi}{\gamma_b^{3/2}\left[ k_{pe}^2-k_\xi^2\right]^{1/2}}, \quad |k_\xi| < k_{pe}.
\end{equation}
A large growth rate occurs for $k_\xi=k_{pe}=1/\delta_{sk}$, which is strongly seeded by the wakes of individual beam electrons, as we observe in Fig. \ref{qeb-fields-mono}.

After an electron bunch forms, the front of the bunch begins to break apart first.  This disintegration occurs because the focusing field in the middle of the beam shifts back, which occurs as bunches form further forward in the beam, leading to a defocusing part of the field overlapping with the front of the bunch.  This process saturates the instabilities.  While saturation occurs at different $s$ values for different parts of the beam, its effect is obvious in the drop in stopping power after the peaks in Fig.\ \ref{dEdx-plots}. The effect of this saturation process in beams large enough to filament is unclear.  Once a bunch breaks apart in one filament, its constituent electrons may move into adjacent filaments, become constituents of new bunches, and cause the stopping power to grow again.

\begin{figure}[tbp]
	\centering
    \includegraphics[width=\linewidth]{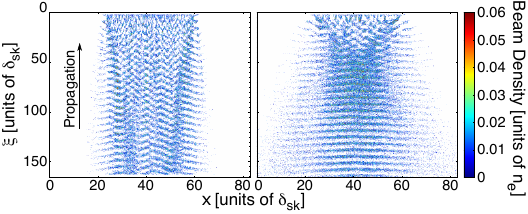}
    \caption{The beam density in a cut plane through the middle of the box at $y=41.515\delta_{sk}$ is plotted for simulation case 4 with $N_b=64$ at $s=200\delta_{sk}$ on the left and $s=400\delta_{sk}$ on the right.\label{qeb-large}}
\end{figure}

The filamentation we observe in Fig.\ \ref{qeb-large} is the primary motivator for the large beam simulations (case 4).  We observe such filamentation when $N_b=8$ and 64.  However, we see no signs of it when $N_b=1/8$ and 1.  We also observe that the stopping power of the two denser beams peaks later than in the small-box simulations, which does not occur with the less dense beams, suggesting that the filamentation is delaying the growth of the instabilities.  The fact that the stopping power for $N_b=64$ peaks with about the same stopping enhancement as in the smaller box also implies that the filamentation may be limiting the enhancement.  Therefore, filamentation may also limit the stopping enhancement in applications where larger beams are used, such as FI.


A second peak in stopping power occurs around $s=1300\delta_{sk}$ for the small monoenergetic beam with $N_b=64$, which is caused by four bunches at the same $\xi$ forming at the very front of the beam.  These bunches are arranged in a square pattern but rotated relative to the initial square cross-section of the beam.  The bunching forces are weakest at the front of the beam, causing those bunches to form later.  This second peak appears down to $N_b=1$ in monoenergetic small-beam simulations with $p_z=10m_ec$, but is caused by a bunch that forms further back in the beam \cite{ellis:dissertation}.  These four bunches are also clear evidence of filamentation, but it does not spread backwards in the beam because the streaming-like instabilities have already saturated there and the beam has diffused transversely.  We do not see the filamentation limitation in other small-beam simulations, likely due to the limited transverse size of the beams.  The beam with a 1 MeV temperature does not have this peak due to the density in the bunch reaching a lower peak level, $\approx0.1n_e$ vs. $\approx0.27n_e$ for the monoenergetic beam. The beam with 1 MeV temperature also does not filament.

\section{Discussion}
The correlated stopping enhancement is more pronounced at lower electron densities than the extreme $10^{26}$ cm$^{-3}$ considered here. Short-pulse lasers generally produce electron beams with $n_b\sim n_c$ at background densities $n_e \lesssim n_c$.  For the typical short-pulse laser wavelength $\lambda=1\ \mu$m, our runs with $N_b=1/8$ correspond to $n_b \approx n_c$ and $n_e\approx 10^5n_c$.  Typical values at the absorption region of $n_b=n_e=n_c$ give extremely high correlation: $N_b= 4.5\times10^6$. The same beam in a solid density target with $n_e=100 n_c$ also has very high correlation: $N_b=4500$. Collisional stopping may be greatly enhanced in these conditions due to correlation effects.

A major open question is whether our enhanced stopping persists over the much larger length and time scales relevant to practical applications.  Past simulations of beams of energetic electrons incident on uniform dense plasmas show that the beams penetrate much deeper than the distance traveled in our simulations.  A Monte Carlo simulation of electron transport including stopping and scattering using a 2D Lagrangian fluid code shows that a beam of monoenergetic 1.5 MeV electrons incident on a 300 g cm$^{-3}$ DT plasma at 5 keV has its peak energy deposition at $\sim$45,000 skin depths \cite{atzeni:stopping}.  Similarly, a hybrid reduced model for relativistic electron beam transport based on the Vlasov-Fokker-Planck equation using an electron beam with mean electron energy of 1.5 MeV incident on a 50 g cm$^{-3}$ hydrogen plasma with a temperature of 1 eV has its peak energy deposition at $\sim$20,000 skin depths \cite{Touati:Reduced}.  There is no indication that beam-plasma instabilities were included in either simulation.  Regardless, they indicate the scale of future simulations that may be required to study the beam transport problem, including beam-plasma-like instabilities.

Our simulations demonstrate that correlation effects can significantly enhance electron beam stopping in HED plasmas.  We observe the stopping power increase to 1200$\times$ the single-electron value for beams with $N_b=64$.  As the beam density decreases, discrete particle-wake interactions become more important, and the fluid approximation breaks down.  All our simulations indicate that beam-plasma-like instabilities lead to an increase in stopping power for $N_b\geq1/8$.  Ignoring the coherent interactions of discrete particle wakes and the related self-focusing, filamentation, and beam-plasma-like instability leaves out important factors in the stopping power.  In particular, because correlated stopping increases with $N_b\propto n_e^{-3/2}$, it may make FI feasible at lower core densities than currently envisaged.  Future work should determine the effects of background temperature, beam divergence, angular scattering, and energy spread, and employ fully electromagnetic codes.

\begin{acknowledgments}

We thank Weiming An, Frank Tsung, and Viktor Decyk for assistance with QuickPIC, OSIRIS, and the UPIC Framework, respectively.  This work performed under the auspices of the U.S.\ Department of Energy by LLNL under Contract DE-AC52-07NA27344 and by UCLA under Grants FG52-09NA29552 and DE-NA0001833.  This work was funded in part by LLNL LDRD projects 09-SI-011, 11-SI-002, and 12-SI-005.  Work by I.\ N.\ Ellis was supported in part by the Lawrence Graduate Scholar Program.

This document was prepared as an account of work sponsored by an agency of the United States government. Neither the United States government nor Lawrence Livermore National Security, LLC, nor any of their employees makes any warranty, expressed or implied, or assumes any legal liability or responsibility for the accuracy, completeness, or usefulness of any information, apparatus, product, or process disclosed, or represents that its use would not infringe privately owned rights. Reference herein to any specific commercial product, process, or service by trade name, trademark, manufacturer, or otherwise does not necessarily constitute or imply its endorsement, recommendation, or favoring by the United States government or Lawrence Livermore National Security, LLC. The views and opinions of authors expressed herein do not necessarily state or reflect those of the United States government or Lawrence Livermore National Security, LLC, and shall not be used for advertising or product endorsement purposes.
\end{acknowledgments}

\appendix

\section{Single-Particle Stopping Power}\label{Appendix-A}
In this appendix, we discuss the basic physics of single-particle stopping power and compare various stopping power formulas with the stopping power measured in QuickPIC using various cell widths.  We provide the formulas here for reference, detailed derivations of which can be found in Ref.\ \cite{ellis:dissertation}.

Collisions in a plasma and the stopping power of energetic particles can be described from two distinct points of view, which we discuss from a classical approach. In one, Lenard-Balescu, they are viewed as the interaction of particles through the plasma response from test charges moving in straight lines (unperturbed orbits). For energetic electrons, the plasma response is a plasma wave wake and the stopping power is determined from the decelerating electric field at the location of the moving particle. In the other, Landau-Boltzmann, they are viewed as two-body Coulomb interactions. In both cases, the resulting stopping power is $\propto\ln\Lambda$ with $\Lambda\equiv b_{max}/b_{min}$, the ratio of a large length $b_{max}$ to a small length $b_{min}$, that diverges for different reasons. In the first case, $b_{max}$ is well defined as a finite screening length, but $b_{min}$ is not defined and is often chosen to be the scale near where large-angle scattering events might occur. In the latter, $b_{min}$ is well defined as the distance of closest approach, but $b_{max}$ is not well defined and is often chosen as the Debye length.  Although not rigorous, the two views are often ``summed''  together such that $\Lambda$ is the ratio of $b_{max}$ from the wake calculation and $b_{min}$ from the two-body collision calculation. The same result follows from the wake approach and cutting off the integral at $b_{min}$. Combining the wake and two-body views has more merit for the relativistic stopping power, as it is not clear how to include quantum effects in the wake calculation.  

Although PIC calculations do not rigorously include QED effects, they can provide a qualitatively correct behavior for the stopping power and naturally permit a study of mutual -- or correlated -- stopping.  As noted above, the stopping power from the wake of a single particle diverges as $b_{min}$ approaches zero. In most cases, the wake is calculated using the Vlasov equation and solving for the wake potential as an integral in wavenumber space. In order obtain simple expressions, the plasma is then assumed to be cold. If the cold limit is considered first, then the electric field from the wake can be obtained from cold fluid theory.  The plasma-based accelerator community has used this approach to calculate the Green's function response, which can be viewed as the response for a single charged particle moving near the speed of light in a cold plasma \cite{katsouleas:beam, ellis:dissertation}.  The axial electric field for the Green's function for a point charge $q$ moving in the $z>0$ direction is 
 \begin{equation}
E_z (r,z) =-2q\delta_{sk}^{-2} K_0 (r/\delta_{sk})\eta(t-z/c) \cos(\omega_{pe} (t-z/c)),\label{cold_green_function}
\end{equation} 
where $r=(x^2+y^2)^{1/2}$, $K_0$ is the modified Bessel function of the second kind, and $\eta(x)$ is the Heaveside step function. Although this Green's function diverges at $r=0$, the response from a beam or a finite-size particle does not \cite{May:Enhanced}. If a particle has a finite size given by a Gaussian charge density $\rho(r,z)=q/((2\pi)^{3/2} \sigma_r^2\sigma_z) \exp[-r^2/2\sigma_r^2-z^2/2\sigma_z^2]$, then $E_z$ on the particle becomes
\begin{equation}
E_z = -q\delta_{sk}^{-2} \ln\left( \frac{1.12\delta_{sk}}{\sigma_r} \right) \label{wake_stopping}.
\end{equation}
Qualitatively, physical quantum particles have a finite size that scales with the de Broglie wavelength $\lambda_{db}$, so the wakes produced by particles of finite size $\sim\lambda_{db}$ are qualitatively similar to those produced by relativistic electrons. In fact, the expression in Eq.\ \ref{QED-dEdx} in the paper can be obtained from evaluating the wake from classical arguments for a finite-size particle and setting $\sigma_{r}$ to $\lambda_{db}$ in the center of mass frame between the moving charge and a plasma electron \cite{ellis:dissertation}. Furthermore, the divergence of the Green's function cannot be correct for a real classical plasma, so the assumptions of a fluid background and linearized response must break down. We will address this issue in a separate publication and below.

We now present several different formulas for relativistic single-electron stopping power.  For convenience, we write the stopping power as
\begin{equation}
\frac{d\gamma}{ds} = -\frac{e^2}{m_ec^2} \frac{\omega_{pe}^2}{v^2} L_d.
\end{equation}
$L_d$ is often called the ``stopping number," and the $d$ in $L_d$ means ``drag."  Other variables are the same as those used in the paper.  As discussed above, $L_d$ is approximately the logarithm of a ratio of two lengths,
\begin{equation}
L_d = \ln\frac{b_{max}}{b_{min}}.
\end{equation}
In all our formulas, we set $b_{max}=v/\omega_{pe}$, which is the dynamic screening length for a moving charge.

The most basic stopping power is that due to a cold fluid wake.  Using the formula for the electric field given in Eq.\ \ref{cold_green_function}, the stopping power is given by
\begin{align}
	L_d=\frac{\delta_{sk}^2}{e}E_z\left(r\to 0,t-\frac{z}{c}=0\right)&= 2K_0(\delta_{sk}^{-1}r\to0)\frac{1}{2}\nonumber\\
	&\approx \ln\left(\frac{\delta_{sk}}{r\to0}\right),
\end{align}
which clearly diverges.  However, the formula can be used to roughly approximate the stopping power of a finite-size particle in a PIC code with cell width $\Delta$, in which case
\begin{equation}
	L_d\approx \ln\left(\frac{\delta_{sk}}{\Delta}\right).\label{eq-wake-stop}
\end{equation}
We compare this formula with single-particle stopping power measured in QuickPIC in Fig.\ \ref{fig-qp-stop-vcw} below.

The stopping power for a relativistic electron taking into account quantum electrodynamics and a dielectric response is given by \cite{icru37}
\begin{align}
L_d = & \ln\left(\frac{[2(\gamma-1)]^{1/2}{m_ecv}}{\hbar\omega_{pe}}\right)-\ln 2 +\frac{9}{16} \nonumber \\
 & + \frac{\ln 2+1/8}{2\gamma^2}-\frac{\ln 2 + 1/8}{\gamma}.\label{Davies-formula-final}
\end{align}
Eq.\ \ref{QED-dEdx} of the paper uses just the first term of Eq.\ \ref{Davies-formula-final}.

Finally, the classical and relativistic (so-called ``Bohr") stopping power formula uses the classical distance of closest approach for $b_{min}$ and is given by \cite{Frenkel:Energy, Prentice:Collective, jackson:classical}
\begin{equation}
L_d = \ln\left(\frac{[(2(\gamma-1)]^{1/2}m_ecv^2}{e^2\omega_{pe}}\right).\label{eq-coll-bohr}
\end{equation}

We compare $\Lambda$ in different models, and discuss the role of PIC particle size.  The physics to bear in mind for our problem is that the classical $b_{min}$ (distance of closest approach for binary collisions) is much smaller than the quantum one (de Broglie wavelength), which, in turn, is much smaller than the PIC spatial grid sizes (PIC particle size) that are feasible on current computers.  The classical result for a relativistic electron is $\Lambda^{cl} \approx \beta\delta_{sk} / b_{min}^{bc} = \Lambda^{qm} \hbar v/e^2$ with $b_{min}^{bc} = e^2/[[2(\gamma-1)]^{1/2}m_ecv]$. We estimate the PIC cell size $\Delta$ imposes $b_{min}^{PIC}\approx \max(b_{min}^{bc},\Delta)$. For our simulations, $\Delta=32,300b_{min}^{bc}$, so $\Lambda^{PIC}_1 = (136/32,300)\Lambda^{qm} = 0.00422 \Lambda^{qm}$, or $\ln\Lambda^{PIC}_1 = \ln\Lambda^{qm} - 5.47$. We use the measured value of this difference in Fig.\ \ref{fig-qp-stop-vcw}, not this estimate.  The quantum, single-particle stopping is thus larger than that in our PIC simulations.  The question arises of how much of this additional single-particle stopping would be enhanced in a PIC simulation with much smaller cell size $\sim\lambda_{db}$. In Fig.\ \ref{fig-qp-stop-vcw}, we present results for how the stopping power of a single electron with momentum $p_z=10m_e c$ increases as the cell size (particle size) decreases from our standard value $\Delta_0$ to $\Delta_0/32$. The PIC stopping increases from roughly $0.3$ to $0.5$ of the quantum stopping. Computer limitations prevent us from carrying out correlated stopping simulations using the smaller cell sizes. 

We compare the stopping power given by the formulas with the single-particle stopping power measured in QuickPIC in Fig.\ \ref{fig-qp-stop-vcw} for an electron with $p_z=10m_ec$.  The simulation parameters are listed in Table \ref{table-qp-stopping-params}.  To compare with the stopping power in Eq.\ \ref{eq-wake-stop}, we vary the cell width in the simulations between the initial cell width $\Delta_0$ and $\Delta_0/32$ while keeping the box size constant.

\begin{table}[htbp]
\centering
\begin{tabular}{ |l|l| }
  \hline
  \multicolumn{2}{|c|}{QuickPIC Simulation Parameters} \\
  \hline
  $n_e$ & $10^{26}$ cm${}^{-3}$ \\
  $T_e$ & 0 eV \\
  Interpolation & Linear\\
  $ds$ & $\delta_{sk}$ for all cell widths\\
  Box Dimensions & $ 10.3787\delta_{sk}[1 \times 1 \times 1/32]$\\
  Box Cells & $ 256 \times 256 \times 8 $ for $\Delta_0=0.0405\delta_{sk}$ \\
  \hline
\end{tabular}
\caption{The parameters for the QuickPIC stopping power simulations.\label{table-qp-stopping-params}}
\end{table}

\begin{figure}[tbp]
\centering
\includegraphics[width=\linewidth]{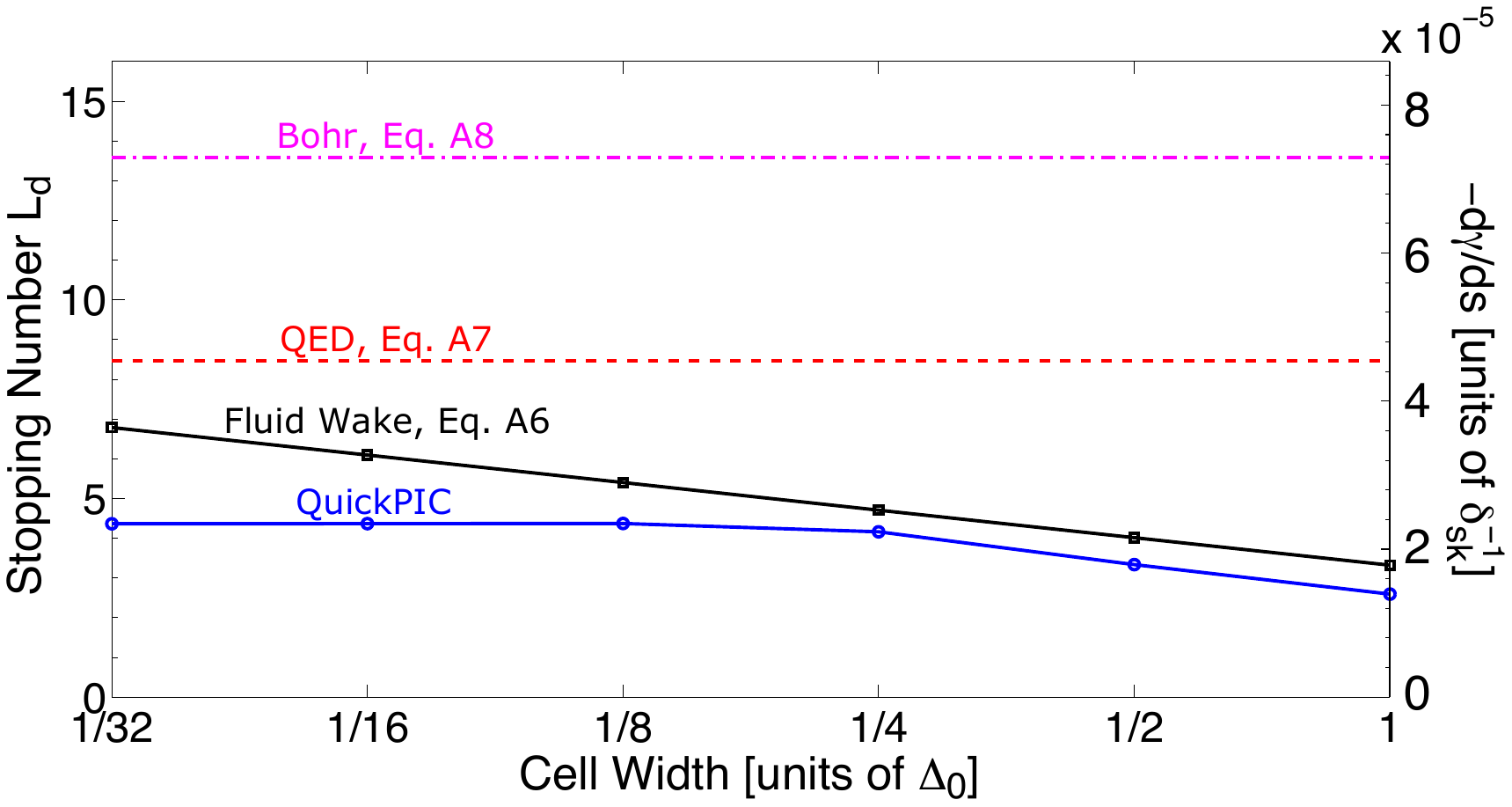}
\caption{The stopping power for an electron with $p_z=10m_ec$ as measured in the QuickPIC simulations across various cell widths and calculated using Bohr, QED, and fluid wake formulas.}\label{fig-qp-stop-vcw}
\end{figure}

The stopping power given by Eq.\ \ref{eq-wake-stop} agrees well with the stopping power measured in QuickPIC until $\Delta<\Delta_0/4$, after which it diverges.  It is possible that taking into account the particle shape when calculating the stopping power would produce better agreement.  The saturation of the stopping power as $\Delta/\Delta_0\to0$ is caused by test electron passing between plasma particles.  As we move the test charge closer to a plasma particle transversely, the stopping power increases.  Therefore, a more accurate simulation of stopping power will require a warm plasma, allowing for random encounters between the test electron and plasma particles.  However, QuickPIC does not currently allow for these simulations due to the numerical instability mentioned in the paper.

Our simulations in the paper simply use a cell width $\Delta=\Delta_0.$  As discussed in the paper, we then use the difference between the single-particle stopping measured in the simulation (using $p_z=20m_ec$) and the QED stopping to bound the enhancement from correlated stopping.  Future research should study the change in beam evolution with cell width, which may lead to a greater understanding of how single-particle stopping power is enhanced by correlation effects.

\section{The Effect of Plasma Temperature on Particle Wakes}\label{Appendix-B}
\begin{figure}[tbp]
\centering
\includegraphics[width=\linewidth]{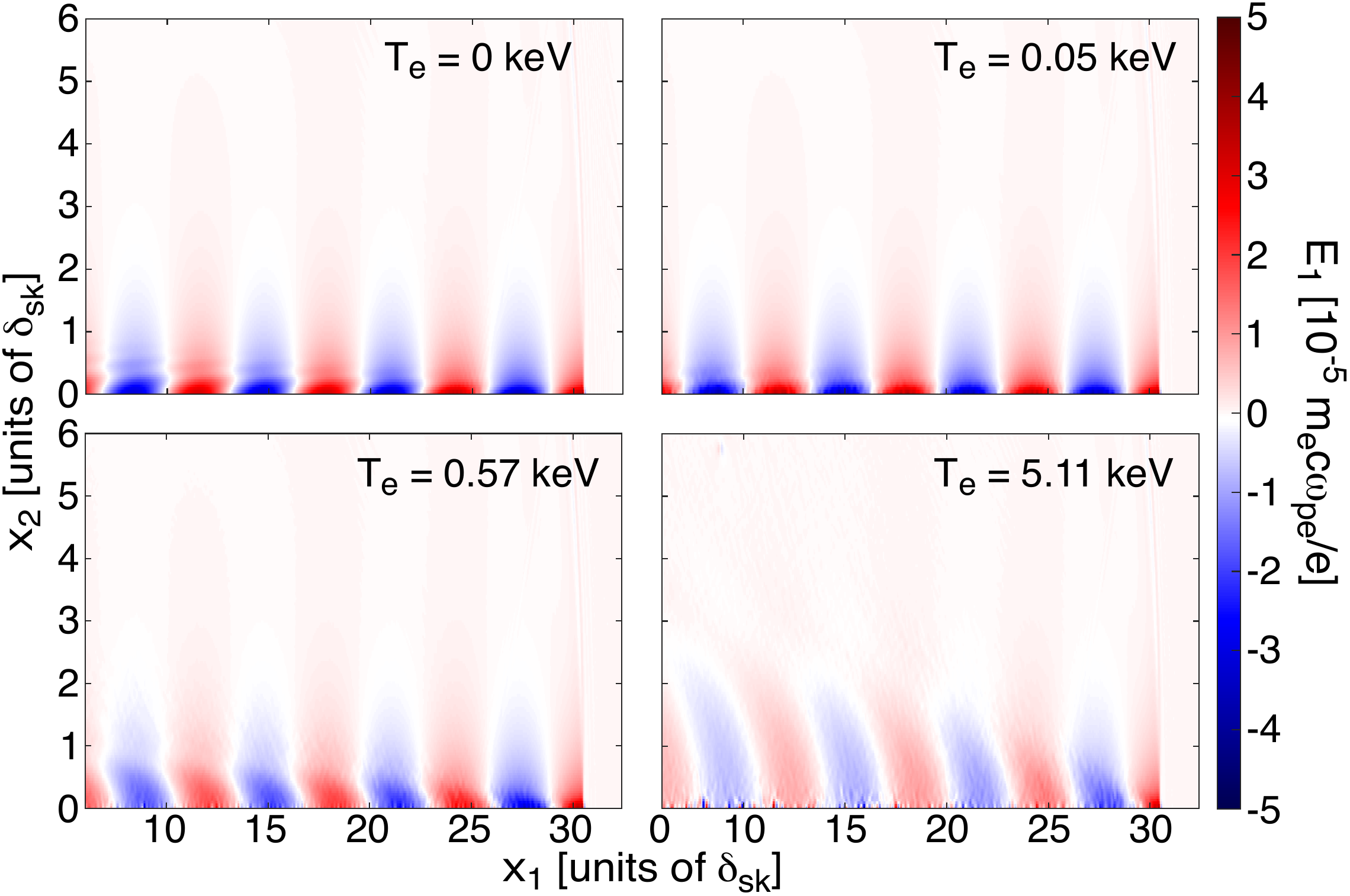}
\caption{The wake from an electron in the quasi-3D version of OSIRIS at various plasma temperatures.  The cell widths in $x_1$ and $x_2$ and plasma density are as in the simulations in the paper.  The parameters work out so that there is approximately 1 PIC particle per real electron.  The test charge is a single electron with $p=20m_ec$ in the $x_1$ direction.}\label{fig-ez-1-part-wake-2d}
\end{figure}

\begin{figure}[tbp]
\centering
\includegraphics[width=\linewidth]{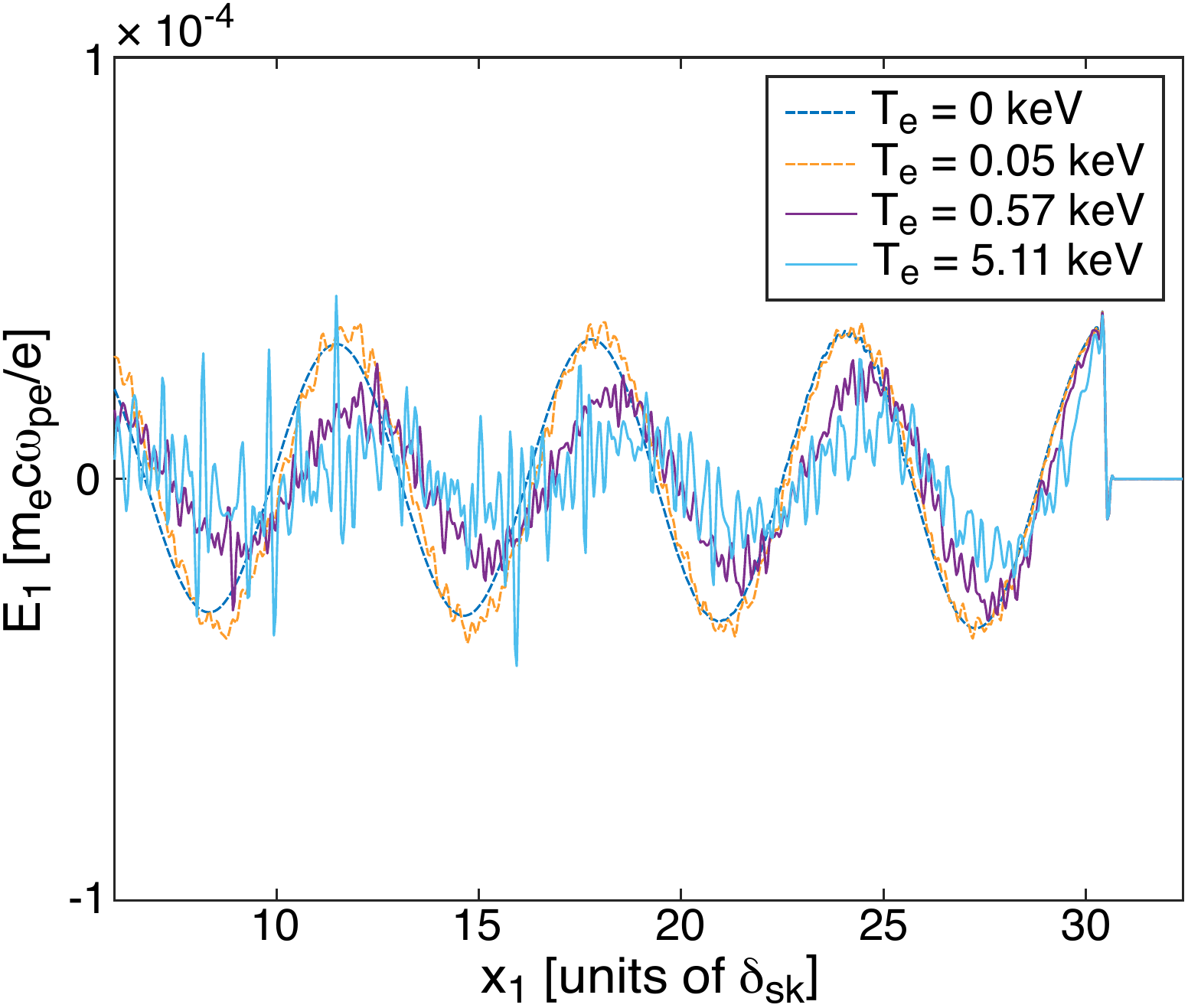}
\caption{The lineout of the electric field of the wake in the $x_1$ direction along the $x_1$ axis for the same runs as in Fig.\ \ref{fig-ez-1-part-wake-2d}.}\label{fig-ez-1-part-wake-lineout}
\end{figure}

To illustrate that the use of cold plasma to study the stopping power is meaningful  we take advantage of  recent progress in simulation capability. Maxwell solvers in standard PIC codes lead to spurious errors in the fields that surround relativistic particles. Recently, a detailed analysis of these showed that they arise due to numerical \v{C}erenkov radiation and aliasing effects \cite{Xu:OnNumerical}. This analysis also indicated how to create a field solver that could mitigate these effects and it was implemented into our code OSIRIS, including a quasi-3D version \cite{Davidson:Implementation}. Using these recent improvements, we simulated the wakes created by a single electron in plasmas with different temperatures.  We used the subtraction technique \cite{decyk:simulation, ellis:dissertation}, in which two identical runs, one with the test charge and one without, and the same random number generator seed for the plasma particles, are conducted. The results from the two simulations are then subtracted, which makes the wake from the test charge clearly visible even though it is below the noise level of the simulation. We show these results in Figs.\ \ref{fig-ez-1-part-wake-2d} and \ref{fig-ez-1-part-wake-lineout}. They clearly show that the field at the location of the particle is nearly identical for all temperatures in the range from 0 to 5 keV.  Furthermore, even after four oscillations, the wakes are clearly present. This indicates that the wake from one electron will persist and influence the trajectories of those behind it in warm plasmas in a qualitatively and quantitatively similar manner as in a cold plasma. These simulations show that wakes in warm plasmas can also spread.

\bibliography{correlated_pre}

\end{document}